\documentclass[%
 reprint,
superscriptaddress,
twocolumn,
 amsmath,amssymb,
 aps,
prb,
]{revtex4-1}

\usepackage[colorlinks=true,linkcolor=blue,urlcolor=blue,citecolor=blue]{hyperref}
\usepackage{graphicx}
\usepackage{dcolumn}
\usepackage{bm}
\usepackage{multirow}
\usepackage{float}
\usepackage{setspace}

\begin{document}


\title{Contrasting electronic states of RuI$_{3}$ and RuCl$_{3}$}


\author{Lu Liu}
 \affiliation{Laboratory for Computational Physical Sciences (MOE),
 State Key Laboratory of Surface Physics, and Department of Physics,
  Fudan University, Shanghai 200433, China}
\affiliation{Shanghai Qi Zhi Institute, Shanghai 200232, China}

\author{Ke Yang}
\affiliation{College of Science, University of Shanghai for Science and Technology, Shanghai 200093, China}
 \affiliation{Laboratory for Computational Physical Sciences (MOE),
 State Key Laboratory of Surface Physics, and Department of Physics,
  Fudan University, Shanghai 200433, China}

\author{Guangyu Wang}
 \affiliation{Laboratory for Computational Physical Sciences (MOE),
 State Key Laboratory of Surface Physics, and Department of Physics,
  Fudan University, Shanghai 200433, China}
\affiliation{Shanghai Qi Zhi Institute, Shanghai 200232, China}

\author{Di Lu}
 \affiliation{Laboratory for Computational Physical Sciences (MOE),
 State Key Laboratory of Surface Physics, and Department of Physics,
  Fudan University, Shanghai 200433, China}
\affiliation{Shanghai Qi Zhi Institute, Shanghai 200232, China}

\author{Yaozhenghang Ma}
 \affiliation{Laboratory for Computational Physical Sciences (MOE),
 State Key Laboratory of Surface Physics, and Department of Physics,
  Fudan University, Shanghai 200433, China}
\affiliation{Shanghai Qi Zhi Institute, Shanghai 200232, China}

\author{Hua Wu}
\email{Corresponding author. wuh@fudan.edu.cn}
\affiliation{Laboratory for Computational Physical Sciences (MOE),
 State Key Laboratory of Surface Physics, and Department of Physics,
 Fudan University, Shanghai 200433, China}
 \affiliation{Shanghai Qi Zhi Institute, Shanghai 200232, China}
\affiliation{Collaborative Innovation Center of Advanced Microstructures,
 Nanjing 210093, China}

%
%

\date{\today}

\begin{abstract}
The spin-orbital entangled states are of great interest as they hold exotic phases and intriguing properties. Here we use first-principles calculations to investigate the electronic and magnetic properties of RuI$_{3}$ and RuCl$_{3}$ in both bulk and monolayer cases. Our results show that RuI$_{3}$ bulk is a paramagnetic metal, which is in agreement with recent experiments. We find that the Ru$^{3+}$ ion of RuI$_{3}$ is in the spin-orbital entangled $j_{\rm eff}=\frac{1}{2}$ state. More interestingly, a metal-insulator transition occurs from RuI$_{3}$ bulk to monolayer, and this is mainly due to the band narrowing with the decreasing lattice dimensionality and to the Ru-I hybridization altered by the I $5p$ spin-orbit coupling. In contrast, RuCl$_{3}$ bulk and monolayer both show Mott-insulating behavior, the Ru$^{3+}$ ion is in the formal $S=\frac{1}{2}$ and $L=1$ state with a large in-plane orbital moment, and this result well explains the experimental large effective magnetic moment of RuCl$_{3}$ and the strong in-plane magnetization. The present work demonstrates the contrasting spin-orbital states and the varying properties of RuI$_{3}$ and RuCl$_{3}$.
\end{abstract}

\maketitle


\section{Introduction}
A wide variety of degrees of freedom, such as crystal field, electron correlation, and spin-orbit coupling (SOC), yield intriguing electronic structures and offer appealing opportunities for novel phenomena and rich properties. In particular, spin-orbital entangled states introduced by the SOC effect have been a hot topic and attracted a vast range of interests in the field of superconductivity, topological phases, quantum spin liquid, and exotic magnetism\cite{2014_annu,superc_2017science,Banerjee_2017science,kasahara_2018nature,Yokoi_2021science}. Among them, the noted $j_{\rm eff}=\frac{1}{2}$ state within the $d$ $t_{2g}$ subshell was first proposed to account for the Mott insulating behavior of Sr$_{2}$IrO$_{4}$\cite{Kim_2008PRL}, and it is now widely used for the $4d$ and $5d$ transition metal compounds with significant SOC, to interpret their exotic electronic and magnetic properties. In addition, the spin-orbital entangled $j_{\rm eff}=\frac{1}{2}$ pseudospin is suggested to accommodate bond dependent interaction in honeycomb lattice\cite{Khaliullin_2005,Khaliullin_2009PRL,kitaevreview_2019}, such as RuCl$_{3}$\cite{Banerjee_2017science, Do_2017natphys}, Na$_{2}$IrO$_{3}$\cite{Na2IrO3_2015natphys}, and Na$_{3}$Co$_{2}$SbO$_{6}$\cite{3Dkitaev_2020PRL}, and it is extensively studied in the realization of the Kitaev model. Such novel spin-orbital states, driven by the delicate interplay of various degrees of freedom, provide room for exploring fundamental physics and potential applications.

As a potential candidate to realize a quantum spin liquid state under application of specific magnetic field strengths and directions, RuCl$_{3}$ is a quasi two dimensional (2D) material, in which the honeycomb layers are coupled by weak van der Waals (vdW) interaction, providing a high possibility to be cleaved into monolayer form with possibly exotic phases. Experimental results reveal that RuCl$_{3}$ bulk is a Mott insulator with planar zigzag antiferromagnetic (AFM) order of T$_{\rm N} \sim$ 7-14 K\cite{Majumder_2015PRB,Sears_2015PRB,Johnson_2015PRB,Kubota_2016PRB,banerjee_2016nm,Cao_2016PRB,Park_2016arxiv,Banerjee_2017science,Do_2017natphys,Sinn_2016}. As a close analog to RuCl$_{3}$, RuI$_{3}$ bulk has been synthesized very recently\cite{Cava_2022am,Nawa_2021JPSJ}. Albeit the same honeycomb lattice and the Ru$^{3+}$ $4d^{5}$ state, RuI$_{3}$ is a paramagnetic (PM) metal\cite{Cava_2022am,Nawa_2021JPSJ}, showing contrasting electronic and magnetic properties from RuCl$_{3}$. Theoretical studies suggest that strong Ru $4d$-I $5p$ hybridization and weak correlation effect account for the measured metallic behavior\cite{Zhang_2022PRB}. It was also proposed that RuI$_{3}$ bulk is a bad metal and is on the verge of the metal-insulator transition\cite{Kaib_2022arx}. The contrasting electronic and magnetic behavior of bulk RuI$_{3}$ and RuCl$_{3}$ stimulate us to study their electronic states, particularly the spin-orbital states out of the intricate interplay of orbital hybridization, crystal field, electron correlation, and SOC.

In this work, using first-principles calculations, we investigate the electronic and magnetic properties of RuI$_{3}$ and RuCl$_{3}$ both in bulk and monolayer forms. Our results show that RuI$_{3}$ is in the PM state due to the strong Ru $4d$-I $5p$ hybridization which suppresses the local Ru $4d$ Hund exchange. These results agree with the experimental and theoretical PM and bad-metallic behavior of RuI$_{3}$ bulk.~\cite{Cava_2022am,Nawa_2021JPSJ,Zhang_2022PRB,Kaib_2022arx} We find that the Ru$^{3+}$ $4d^{5}$ ion is in the $j_{\rm eff}=\frac{1}{2}$ state, and predict that RuI$_{3}$ undergoes a metal-insulator transition from bulk to monolayer. Moreover, we find that the Ru$^{3+}$ ion in RuCl$_{3}$ bulk and monolayer is in the $S=\frac{1}{2}$ and $L=1$ state with a large in-plane orbital moment. This result well accounts for experimental observations of the large effective magnetic moment and strong in-plane magnetization in RuCl$_{3}$.~\cite{Majumder_2015PRB,Sears_2015PRB,banerjee_2016nm,Banerjee_2017science} Thus, we have identified the contrasting electronic structures and magnetic properties for RuI$_{3}$ and RuCl$_{3}$.

\begin{figure}[t]
  \centering
\includegraphics[width=8.5cm]{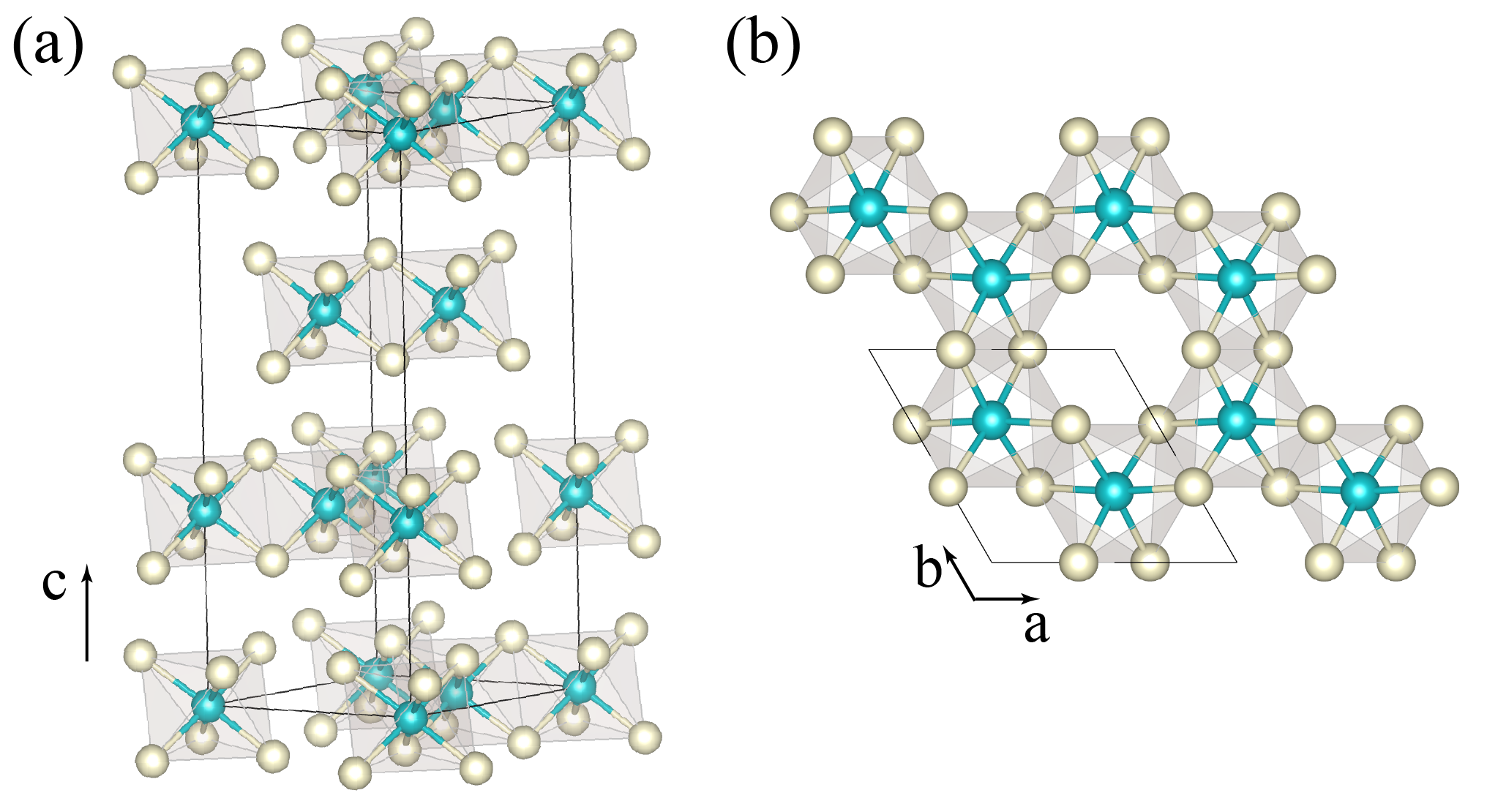}
  \caption{The crystal structure of RuI$_{3}$ bulk (a)  and monolayer (b): Ru (I) atoms are indicated by blue (yellow) balls. }
  \label{fig:1}
\end{figure}

\section{Computational Details}
Density functional theory calculations are carried out using the full-potential augmented plane wave plus local orbital code (WIEN2K)\cite{WIEN2K}. Here we adopt the experimental $R\bar{3}$\cite{Cava_2022am} structure for RuI$_{3}$ bulk and the $C2/m$\cite{Johnson_2015PRB,Cao_2016PRB,banerjee_2016nm} one for RuCl$_{3}$ bulk, both of which have the common edge-sharing octahedra (RuI$_{6}$ vs RuCl$_{6}$) forming a planar honeycomb lattice but have the different stacking orderings. As an example, we show the $R\bar{3}$ crystal structure of RuI$_{3}$ bulk in Fig. \ref{fig:1}. The optimized lattice constants of RuI$_{3}$ bulk (monolayer) are $a$=$b$=6.861 (6.667) \AA~and $c$=18.839 \AA, which are almost the same (within $\sim$1$\%$) as the experimental ones of $a$=$b$=6.791 \AA~and $c$=19.026 \AA\cite{Cava_2022am}. And for RuCl$_{3}$ bulk (monolayer), the optimized parameters are $a$=5.875 (5.769) \AA, $b$=10.167 (10.023) \AA~and $c$=5.911 \AA, and they also agree well (within $\sim$1.7$\%$) with the experimental ones of $a$=5.976 \AA, $b$=10.342 \AA~and $c$=6.013 \AA\cite{Johnson_2015PRB}. We also consider the $P3_{1}12$\cite{banerjee_2016nm} and $R\bar{3}$\cite{Park_2016arxiv} structures for RuCl$_{3}$ bulk, and find that they yield very similar results with the $C2/m$ one; see the Supplemental Material (SM)\cite{[{See Supplemental Material for (i) RuCl$_{3}$ bulk in $P3_{1}12$ and $R\bar{3}$ structures, (ii) LSDA+SOC+U and GGA+SOC+U for RuI$_{3}$, (iii) LSDA+SOC+U and GGA+SOC+U for RuCl$_{3}$, and (iv) larger U values for RuCl$_{3}$ bulk}]SM}. The muffin-tin sphere radii are chosen to be 2.2, 2.4, and 2.1 bohrs for Ru, I, and Cl atoms, respectively. The cutoff energy of 14 Ry is used for plane wave expansion. The integration over the first Brillouin zone is performed using an 11$\times$11$\times$3 (9$\times$5$\times$9) and 11$\times$11$\times$1 (9$\times$5$\times$1) k-mesh for RuI$_{3}$ (RuCl$_{3}$) bulk and monolayer, respectively. The electron correlation effect is included by using the local spin density approximation plus Hubbard $U$ (LSDA+$U$) method\cite{Anisimov_1993}. $U=$ 2 eV and the Hund exchange $J_{\rm H}=$ 0.5 eV are used for the Ru $4d$ electrons \cite{Zhang_2022PRB,Kaib_2022arx}. The SOC is included by the second variational method with scalar relativistic wave functions. We also perform calculations using the generalized gradient approximation (GGA)\cite{PBE} for the exchange correlation functional, and find that the GGA gives quite similar results to the LDA for both RuI$_{3}$ and RuCl$_{3}$ in bulk and monolayer; see the SM\cite{SM}.

\begin{figure}[t]
  \centering
\includegraphics[width=8.5cm]{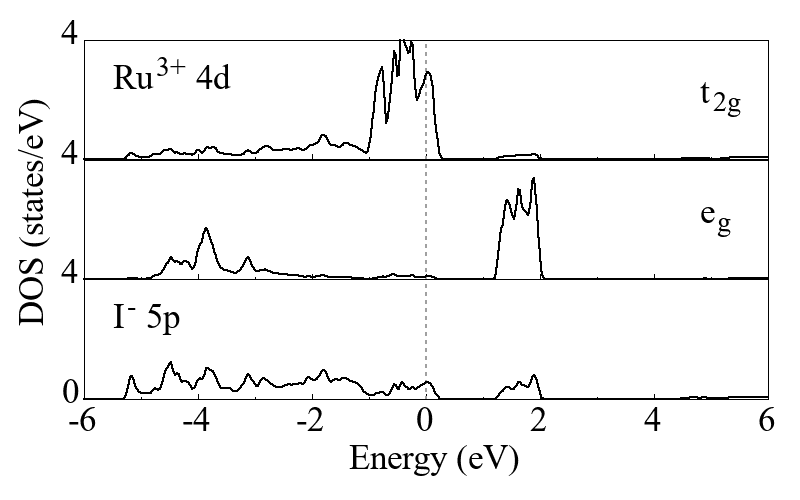}
  \caption{The DOS result of RuI$_{3}$ bulk by LDA. The Fermi level is set at the zero energy.}
  \label{fig:2}
\end{figure}

\section{Results}
\subsection*{A. RuI$_{3}$ bulk: $j_{\rm eff}=\frac{1}{2}$ PM metallic state}

We first investigate RuI$_{3}$ bulk to understand the Ru$^{3+}$ $4d^{5}$ state and the electronic and magnetic properties. To estimate the crystal field effect, we carry out spin-restricted LDA calculations. As shown in Fig. \ref{fig:2}, the large t$_{2g}$-e$_{g}$ splitting of about 2.0 eV makes the unoccupied e$_{g}$ states lie above the Fermi level by about 1.8 eV, and the partially occupied t$_{2g}$ states cross the Fermi level, showing the formal Ru$^{3+}$ $t_{2g}^{5}$ configuration. Besides, the spatially extended I $5p$ orbitals have strong hybridization with Ru $4d$ ones, making the Ru $4d$-I $5p$ states below the Fermi level distribute over the large energy range more than 5 eV. It is worth noting that the $t_{2g}^{5}$ state may carry an unquenched orbital angular momentum and the SOC strength is normally significant for Ru $4d$ electrons. Therefore, the SOC effect of the Ru $4d$ states (and of the I $5p$ orbitals) would be of concern.

\begin{figure}[t]
  \centering
\includegraphics[width=8.5cm]{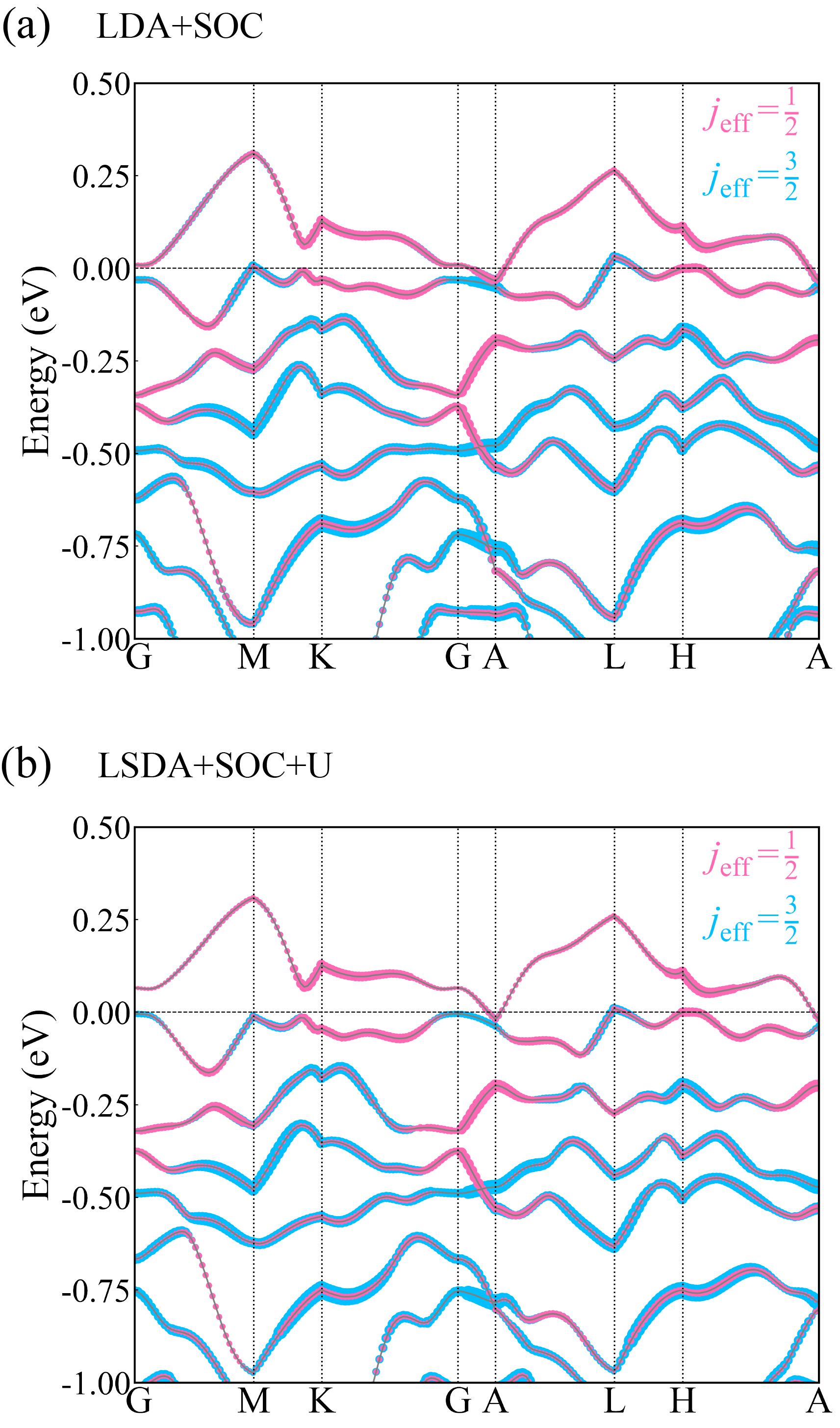}
  \caption{The $j_{\rm eff}=\frac{1}{2}$ (represented by pink curves) and $j_{\rm eff}=\frac{3}{2}$ (blue curves) decomposed band structures of (a) LDA+SOC and (b) LSDA+SOC+U for RuI$_{3}$ bulk in the nonmagnetic state. The Fermi level is set at the zero energy.}
  \label{fig:3}
\end{figure}

We now carry out LDA+SOC calculations to see the SOC effect in RuI$_{3}$. As shown in Fig. \ref{fig:3}(a), in the presence of SOC, t$_{2g}$ orbitals split into a lower $j_{\rm eff}=\frac{3}{2}$ quartet and a higher $j_{\rm eff}=\frac{1}{2}$ doublet. For the t$_{2g}^{5}$ configuration, four electrons fully occupy the $j_{\rm eff}=\frac{3}{2}$ states, and the remaining one gives the half-filled $j_{\rm eff}=\frac{1}{2}$ band across the Fermi level. While there are mixtures of $j_{\rm eff}=\frac{1}{2}$ and $j_{\rm eff}=\frac{3}{2}$, which are contributed by hybridization of Ru $4d$ and I $5p$ orbitals and local distortions away from perfect cubic conditions, those bands near the Fermi level are predominantly from the $j_{\rm eff}=\frac{1}{2}$ states, and the bands in the range of 0.2-1.0 eV below Fermi level mostly consist of $j_{\rm eff}=\frac{3}{2}$ states. Those separate bands via the splitting of $j_{\rm eff}=\frac{1}{2}$ and $j_{\rm eff}=\frac{3}{2}$ show the notable SOC effect of the Ru$^{3+}$ t$_{2g}^{5}$ electrons.

Then we perform LSDA+SOC and LSDA+SOC+$U$ calculations to check the possible effect of Hund's exchange coupling and the moderate electron correlation. For RuI$_{3}$ bulk, we test a ferromagnetic state using LSDA+SOC, but it turns out to be unstable and converges to the same nonmagnetic state as the above LDA+SOC solution. In this sense, the Hund exchange is ineffective in RuI$_{3}$ bulk, which seems to be suppressed by the strong Ru $4d$-I $5p$ band hybridizations. Then we probe the electron correlation effect using LSDA+SOC+$U$ calculations. We consider the nonmagnetic, ferromagnetic, and zigzag AFM configurations. All these states turn out to have a very close total energy within 0.4 meV/f.u., and the ferromagnetic and the zigzag AFM states have only a small magnetic moment of about 0.1 $\mu_{\rm B}$/Ru. Those results indicate that RuI$_{3}$ bulk is nonmagnetic rather than magnetic, and the electron correlation effect is insignificant by a comparison of the band structures shown in Figs. 3(a) and 3(b). The present nonmagnetic and metallic solution is largely ascribed to the strong Ru $4d$-I $5p$ hybridizations and thus the delocalization behavior of the Ru $4d$ electrons. Therefore, our results well reproduce the experimental PM and bad-metallic behavior of RuI$_{3}$ bulk\cite{Cava_2022am,Nawa_2021JPSJ} which is on the verge of the metal-insulator transition~\cite{Kaib_2022arx}, and show that RuI$_{3}$ is in the spin-orbital entangled $j_{\rm eff}=\frac{1}{2}$ state. 

\begin{figure}[t]
  \centering
\includegraphics[width=8.5cm]{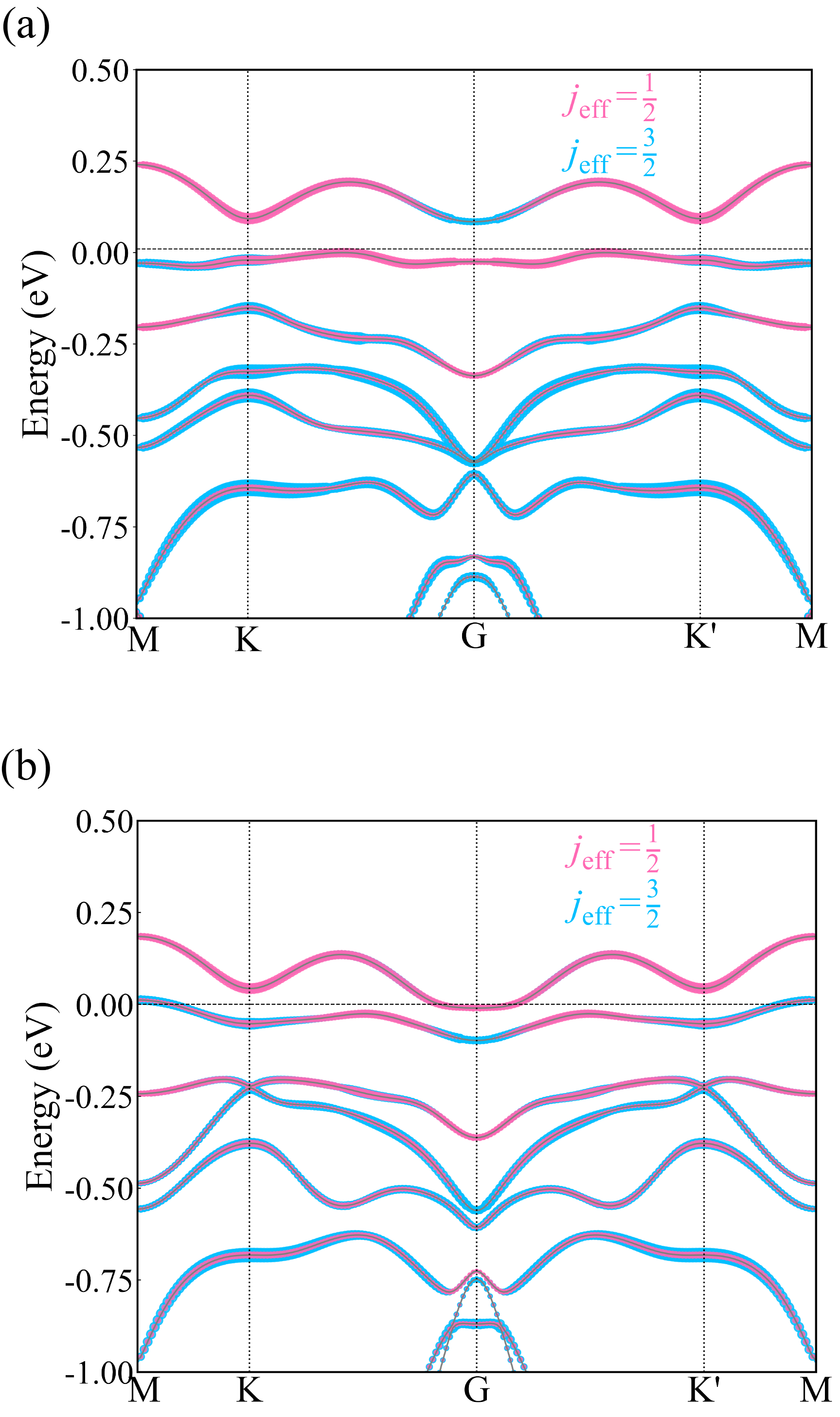}
  \caption{The $j_{\rm eff}=\frac{1}{2}$ (represented by pink curves) and $j_{\rm eff}=\frac{3}{2}$ (blue curves) decomposed band structures of LDA+SOC for RuI$_{3}$ monolayer. (a) Both the Ru and I SOC are included. (b) The I SOC is inactive and only Ru SOC is active. The Fermi level is set at the zero energy.}
  \label{fig:4}
\end{figure}

\subsection*{B. RuI$_{3}$ monolayer: $j_{\rm eff}=\frac{1}{2}$ PM insulating state}

\begin{figure}[t]
  \centering
\includegraphics[width=8.5cm]{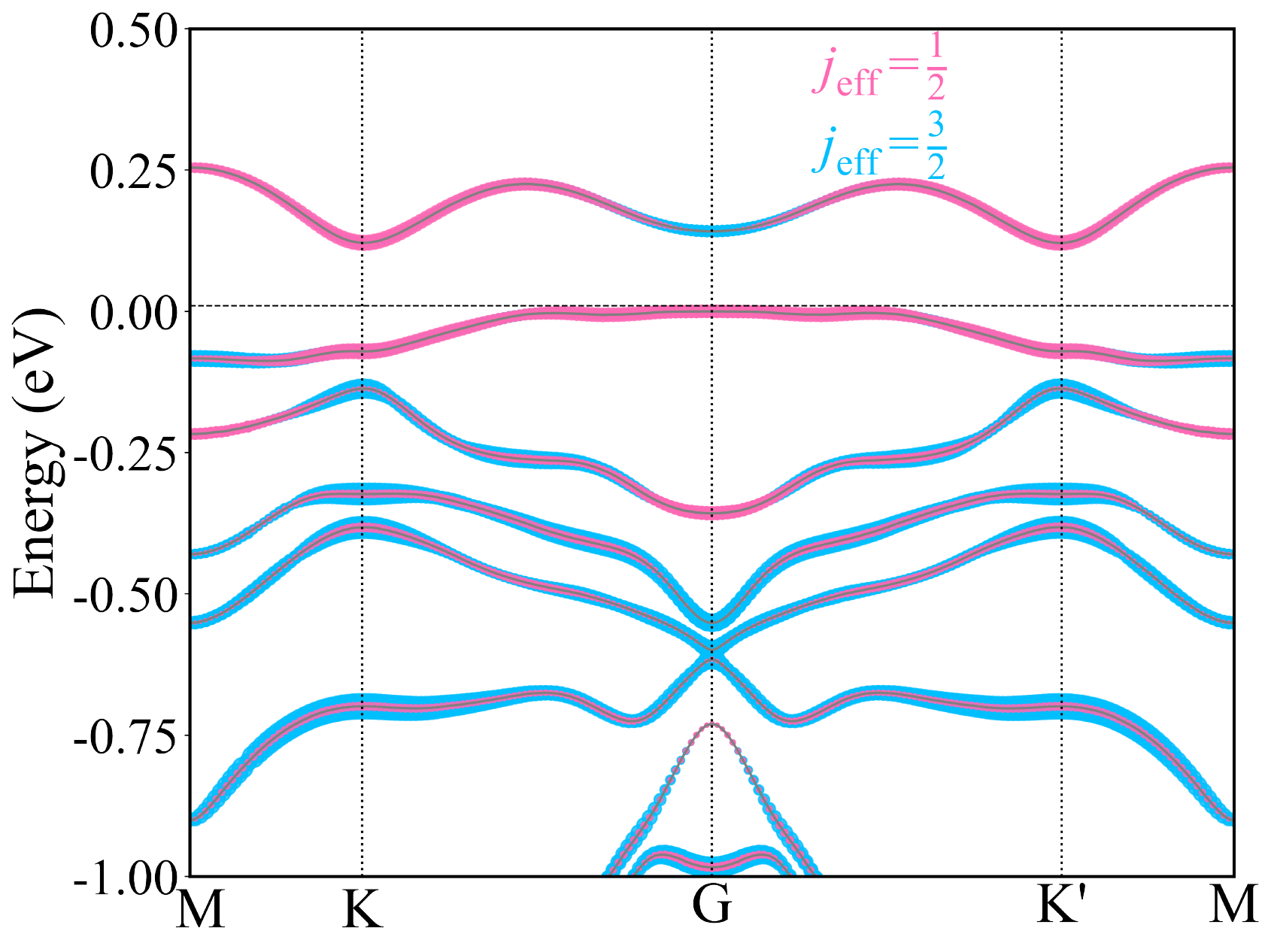}
  \caption{The $j_{\rm eff}=\frac{1}{2}$ (represented by pink curves) and $j_{\rm eff}=\frac{3}{2}$ (blue curves) decomposed band structure of LSDA+SOC+$U$ for RuI$_{3}$ monolayer. The Fermi level is set at the zero energy.}
  \label{fig:5}
\end{figure}

As a quasi 2D material, RuI$_{3}$ bulk may be cleaved into the monolayer form.  Using the DFT$+$vdW correction\cite{vdw}, the cleavage energy is calculated to be 0.24 J/m$^{2}$, and it is even smaller than that~\cite{CrI_cleave} of 0.3 J/m$^{2}$ for CrI$_{3}$ which is already successfully cleaved by mechanical exfoliation\cite{Huang_2017}. Thus, we are now motivated to study the electronic and magnetic properties of the RuI$_{3}$ monolayer. Again, both the LDA+SOC and LSDA+SOC calculations give the same nonmagnetic solution. In contrast to the above nonmagnetic metallic state for RuI$_{3}$ bulk [Fig. 3(a)], the present nonmagnetic solution for the RuI$_{3}$ monolayer has now an insulating gap as seen in Fig. 4(a). In this sense, RuI$_{3}$ undergoes an interesting metal-insulator transition from bulk to monolayer. Besides the common reason of the band narrowing for gap opening due to the reduced dimensionality in the monolayer, the SOC effect of the I $5p$ orbital is found to be an important reason via the strong Ru $4d$-I $5p$ hybridization. By a comparison between Figs. 4(a) and 4(b), we find that when the I $5p$ SOC is switched off in the LDA+SOC calculations, the two bands of the $j_{\rm eff}=\frac{1}{2}$ subset around the Fermi level change a lot in energy dispersion and band shift, and they restore the metallic behavior. Note that for a typical $j_{\rm eff}=\frac{1}{2}$ magnetic Mott insulator like Sr$_2$IrO$_4$, the half-filled $j_{\rm eff}=\frac{1}{2}$ band undergoes a split and opens an insulating gap by a Hubbard $U$. In strong contrast, here for the nonmagnetic insulating RuI$_{3}$ monolayer, the gap opening is not due to a Hubbard $U$, but to the band narrowing in the reduced dimensionality and to the band tuning of  the strong Ru $4d$-I $5p$ hybridization by the I $5p$ SOC effect.

Now we check the possible effects of the Hubbard $U$ on the electronic and magnetic structures of the RuI$_{3}$ monolayer. By carrying out LSDA+SOC+$U$ calculations, we can stabilize the nonmagnetic, ferromagnetic, and zigzag AFM solutions, respectively. Our results show that all these three solutions are in the $j_{\rm eff}=\frac{1}{2}$ state, and that the ferromagnetic (zigzag AFM) state has the local spin/orbital moment of 0.50/0.22 (0.42/0.44) $\mu_{\rm B}$/Ru. It is important to note that the nonmagnetic state is most stable and it has a lower total energy than the zigzag AFM and ferromagnetic states by 2.7 and 15.1 meV/f.u., respectively. The nonmagnetic insulating gap is slightly increased by the Hubbard $U$; see Figs. 4(a) and (5). This nonmagnetic $j_{\rm eff}=\frac{1}{2}$ insulating state is much different from the typical magnetic $j_{\rm eff}=\frac{1}{2}$ Mott-insulating state, e.g., in Sr$_{2}$IrO$_{4}$\cite{Kim_2008PRL}. The latter has a formal local spin (orbital) moment of 0.33 (0.67) $\mu_{\rm B}$, and its Mott gap is opened by the correlation effect in the half-filled $j_{\rm eff}=\frac{1}{2}$ doublet. For the RuI$_{3}$ monolayer, it is found to be a nonmagnetic $j_{\rm eff}=\frac{1}{2}$ insulator, and its gap opening is mainly due to the band effect of the strong Ru $4d$-I $5p$ hybridization with the significant SOC, rather than the electron correlation effect, as proven in the above calculations. In this respect, the previous theoretical suggestion of a ferromagnetic behavior~\cite{Huang_RuI3,Ersan_RuI3} for the insulating RuI$_{3}$ monolayer seems at odd with the experimental PM behavior of RuI$_{3}$ bulk,\cite{Cava_2022am,Nawa_2021JPSJ} as the weak interlayer vdW interaction normally would not change the intralayer magnetism drastically. Therefore, our present result of the nonmagnetic behavior for the RuI$_{3}$ monolayer would be more reasonable. 

So far, using the first principles calculations, we have well explained the experimental PM metallic state of RuI$_{3}$ bulk~\cite{Cava_2022am,Nawa_2021JPSJ}, and have predicted the interesting nonmagnetic insulating state for the RuI$_{3}$ monolayer. Both the bulk and monolayer RuI$_{3}$ are in the $j_{\rm eff}=\frac{1}{2}$ state due to the strong Ru $4d$ SOC. But an interesting metal-insulator transition occurs from the bulk to monolayer, and it is mainly due to the band narrowing in the reduced dimensionality and to the band tuning of the strong Ru $4d$-I $5p$ hybridization by the I $5p$ SOC effect.

\subsection*{C. RuCl$_{3}$ bulk: $S=\frac{1}{2}$ and $L=1$ Mott-insulating state}

\begin{figure}[t]
  \centering
\includegraphics[width=8.5cm]{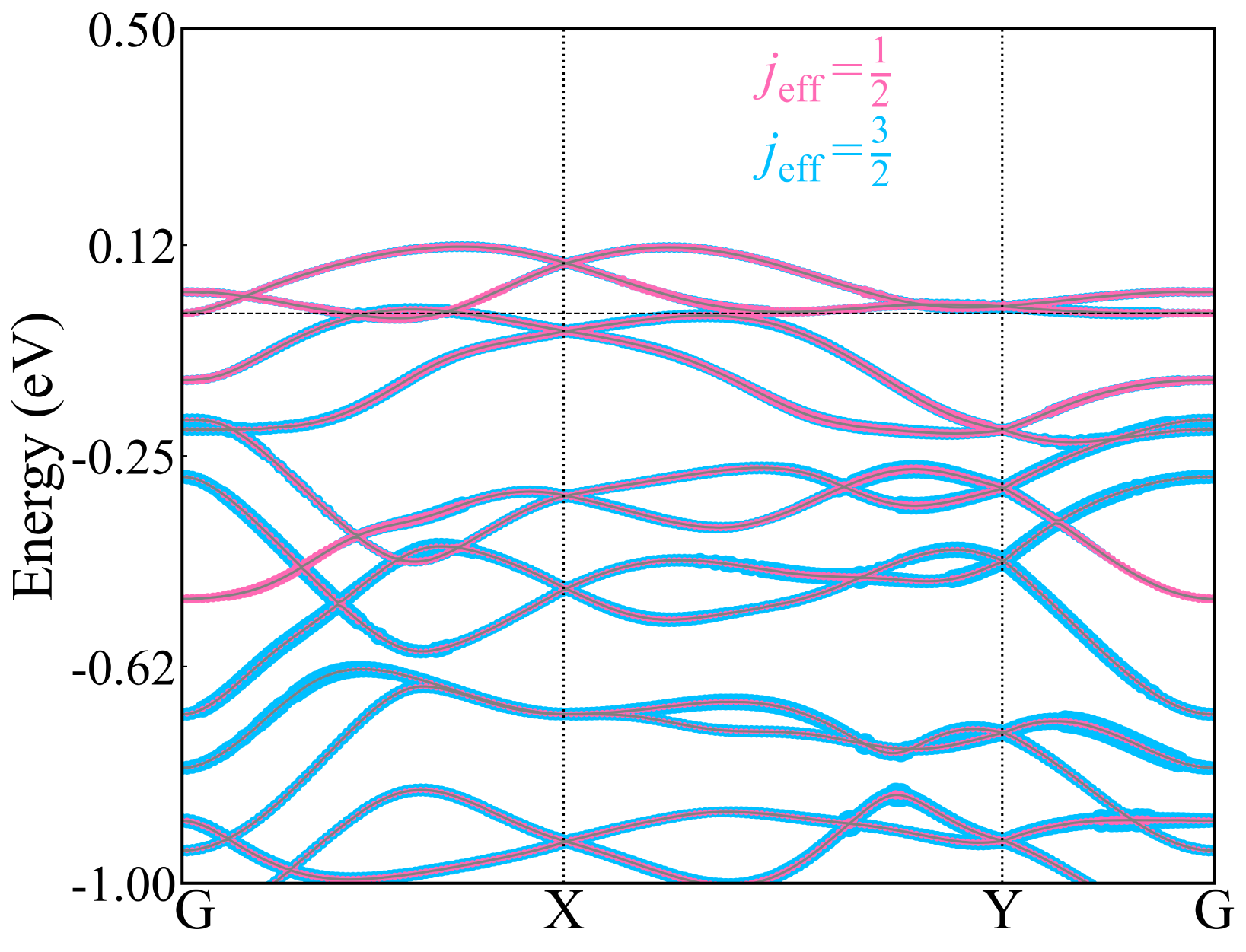}
  \caption{The $j_{\rm eff}=\frac{1}{2}$ (represented by pink curves) and $j_{\rm eff}=\frac{3}{2}$ (blue curves) decomposed band structure of LDA+SOC for RuCl$_{3}$ bulk. The Fermi level is set at the zero energy.}
  \label{fig:6}
\end{figure}

Comparing with RuI$_{3}$, RuCl$_{3}$ has a similar honeycomb lattice and the same Ru$^{3+}$ $4d^{5}$ configuration, but has totally contrasting electronic and magnetic properties, being a Mott insulator with zigzag AFM order and strong in-plane magnetic anisotropy\cite{Majumder_2015PRB,Sears_2015PRB,Johnson_2015PRB,Kubota_2016PRB,banerjee_2016nm,Cao_2016PRB,Park_2016arxiv,Banerjee_2017science,Do_2017natphys}. This implies diverse spin-orbital states through interplay of various degrees of freedom. Hence, we are motivated to investigate the electronic structures of RuCl$_{3}$ bulk. First, in the LDA+SOC framework, the Ru$^{3+}$ ion of RuCl$_{3}$ bulk is in the $j_{\rm eff}=\frac{1}{2}$ state due to the significant SOC effect of Ru $4d$ electrons, and the $j_{\rm eff}=\frac{1}{2}$ bands cross the Fermi level; see Fig. \ref{fig:6}. The $j_{\rm eff}=\frac{1}{2}$ metallic behavior is similar to the situation in RuI$_{3}$ bulk [Fig. \ref{fig:3}(a)]. 

However, when including Hund's coupling, the electronic state undergoes a drastic change: the Ru $4d$ electrons in RuCl$_{3}$ prefer to be spin polarized, in stark contrast with the PM character of RuI$_{3}$. Now the zigzag AFM state of RuCl$_{3}$ bulk with the easy in-plane magnetization (see below) is calculated by LSDA+SOC to be more stable than the nonmagnetic one by 15.6 meV/f.u., and has a local spin moment of 0.59 $\mu_{\rm B}$/Ru. This suggests that the Hund exchange is effective in RuCl$_{3}$, which should arise from the stronger ionic behavior and the weaker covalence effect in RuCl$_{3}$ than in RuI$_{3}$.

To show the collective effects of crystal field, SOC, Hund's coupling, and electronic correlation, we then perform LSDA+SOC+$U$ calculations. Indeed, the zigzag AFM state is more stable than the nonmagnetic one by 117.7 meV/f.u., showing the strong effects of Hund's exchange and electron correlation. In addition to the local spin moment of 0.70 $\mu_{\rm B}$/Ru, the Ru$^{3+}$ ion in RuCl$_{3}$ bulk also has an in-plane orbital moment of 0.47 $\mu_{\rm B}$/Ru. This state with such a large spin and orbital moment then questions the $j_{\rm eff}=\frac{1}{2}$ description in RuCl$_{3}$, as the $j_{\rm eff}=\frac{1}{2}$ state has a spin moment of 0.33 $\mu_{\rm B}$ and orbital moment of 0.67 $\mu_{\rm B}$\cite{Kim_2008PRL}. In fact, the nature of the $j_{\rm eff}=\frac{1}{2}$ basis in RuCl$_{3}$ remains controversial. In experiments, strong magnetic anisotropy is observed and the g factors are estimated to be g$_{ab}\sim$ 2.5 and g${_c}\sim$ 0.4\cite{Kubota_2016PRB}, whereas theoretical studies demonstrate that the pure $j_{\rm eff}=\frac{1}{2}$ state (i.e., no mixture with $j_{\rm eff}=\frac{3}{2}$) has isotropic orbital nature and gives isotropic g factors. Besides, the experimental magnetic moment of $\sim$1.2 $\mu_{\rm B}$/Ru under an in-plane magnetic filed of 60 T is not yet saturated~\cite{Johnson_2015PRB,Kubota_2016PRB} but is already larger than the total magnetic moment of 1 $\mu_{\rm B}$ either in the $j_{\rm eff}=\frac{1}{2}$ state or in the pure $S=\frac{1}{2}$ state. This indicates a large contribution from the orbital moment, and the in-plane moment of $\sim$1.2 $\mu_{\rm B}$/Ru can be explained by the above calculated spin moment of 0.70 $\mu_{\rm B}$ plus the orbital moment of 0.47 $\mu_{\rm B}$. Moreover, the experimental effective magnetic moment ($\mu_{\rm eff}$) of 2.0-2.4 $\mu_{\rm B}$ for in-plane magnetization\cite{Majumder_2015PRB,Sears_2015PRB,banerjee_2016nm,Banerjee_2017science} is also much larger than that of 1.73 $\mu_{\rm B}$ for the $j_{\rm eff}=\frac{1}{2}$ state. Furthermore, theoretical studies suggest that local distortions and energy splitting, e.g., splitting via trigonal crystal field, would alter orbital and spin components of $j_{\rm eff}=\frac{1}{2}$ basis and bring about anisotropic behavior\cite{Kubota_2016PRB,Winter_2017JPCM,Khaliullin_216PRB}. Those experimental and theoretical works indicate that the $j_{\rm eff}=\frac{1}{2}$ picture of RuCl$_{3}$ may need a reconsideration. As seen below, indeed we find that owing to the considerable Hund’s coupling, moderate SOC, and trigonal crystal field splitting, RuCl$_{3}$ bulk is in the $S=\frac{1}{2}$ and $L_{x}=1$ state with in-plane anisotropy rather than the $j_{\rm eff}=\frac{1}{2}$ state.

\begin{figure}[t]
  \centering
\includegraphics[width=8.5cm]{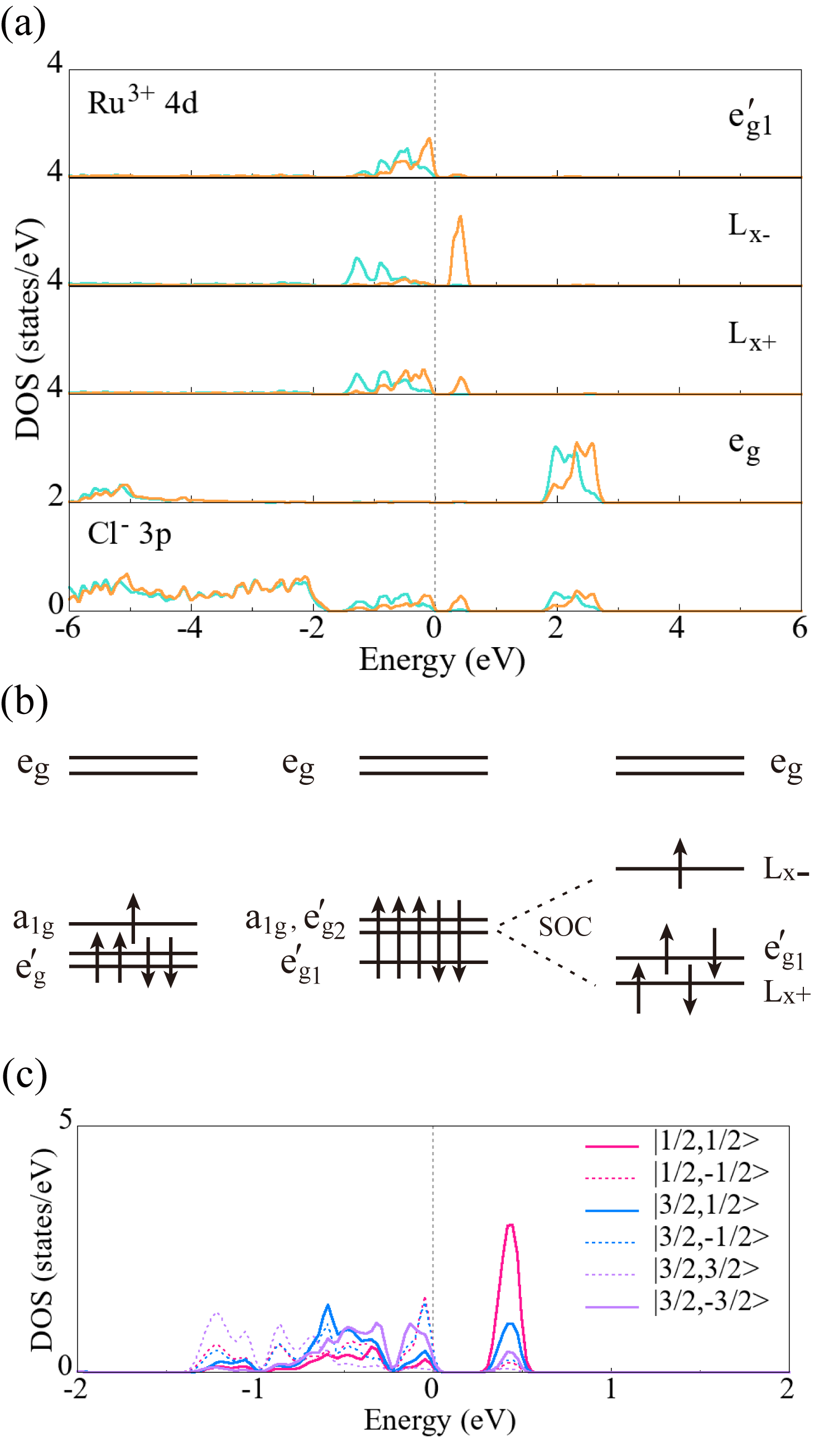}
  \caption{(a) The DOS result of LSDA+SOC+$U$ for RuCl$_{3}$ bulk. The Ru$^{3+}$ ions in zigzag antiferromagnetic order with spins along x axis have each the $t_{2g}$$^{3\uparrow}$$e^{\prime\downarrow}_{g1}$L$_{x+}^{\downarrow}$ configuration. The green (orange) curves refer to up (down) spins. The Fermi level is set at the zero energy. (b) The trigonal crystal field level diagrams, and the SOC mixing of $a_{1g}$ and $e^{\prime}_{g2}$ to yield the L$_{x \pm}$ states. (c) The $t_{2g}$ states in (a) are projected onto the $\left|j_{\rm eff}, m^{x}_{j}\right\rangle$ basis.}
  \label{fig:7}
\end{figure}

Considering the trigonal crystal field, the $t_{2g}$ triplet would split into the $a_{1g}$ singlet and $e'_{g}$ doublet\cite{Khomskii_2001,Yang_2019}. Using the global coordinate system with z axis along the octahedral $\left[ 111 \right]$ direction and y along the $\left[ 1\overline{1}0 \right]$ direction, the wave functions of $t_{2g}$ orbitals can be written as
\begin{equation}
\begin{aligned}
&\left|a_{1 g}\right\rangle=\left|3 z^{2}-r^{2}\right\rangle \\
&\left|e_{g 1}^{\prime}\right\rangle= \sqrt{\frac{2}{3}}\left|x^{2}-y^{2}\right\rangle-\sqrt{\frac{1}{3}}\left|xz\right\rangle \\
&\left|e_{g 2}^{\prime}\right\rangle= \sqrt{\frac{2}{3}}\left|xy\right\rangle+\sqrt{\frac{1}{3}}\left|yz\right\rangle. \\
\end{aligned}
\end{equation}
For the spins aligning in the ab plane, for example, along the x axis, SOC would mix $a_{1g}$ and $e'_{g2}$ orbitals to produce the $L_{x \pm}$ ($L_x=\pm$1) states\cite{lliu_2020,ludi_2022}
\begin{equation}
\begin{aligned}
\left|L_{x}=\pm1\right\rangle= \sqrt{\frac{1}{2}}(i\left|e^{\prime}_{g2}\right\rangle\mp\left|a_{1g}\right\rangle).
\end{aligned}
\end{equation}
Using this basis, the decomposed DOS results are shown in Fig. \ref{fig:7}(a). The Ru$^{3+}$ ion in RuCl$_{3}$ has a moderate exchange splitting of about 0.2 eV, which is somewhat larger than the SOC strength $\lambda$ $\sim$ 0.1 eV for Ru $4d$ electrons\cite{banerjee_2016nm,Winter_2017JPCM,rucl_j_2015PRB,rucl_j_2016prl}. With the assistance of Hund's coupling, the $t_{2g}$$ ^{3\uparrow 2\downarrow}$ configuration is produced. Then the two spin-down electrons occupy the $e'_{g1}$ and $L_{x+}$ states to gain the SOC splitting energy and produce the $S=\frac{1}{2}$ and $L_x=1$ state; see Fig. \ref{fig:7}(a) and (b). A Mott gap is opened by the correlation effect. 

 We also check the $j_{\rm eff}=\frac{1}{2}$ basis by projecting the above $S=\frac{1}{2}$ and $L_x=1$ state (Fig. \ref{fig:7}(a)) onto the $\left|j_{\rm eff}, m^{x}_{j}\right\rangle$ orbitals which have the forms as follows:
\begin{equation}
\begin{aligned}
&\left|\frac{1}{2}, \frac{1}{2}\right\rangle=-\sqrt{\frac{2}{3}}\left|L_{x}=-1,\downarrow\right\rangle+\sqrt{\frac{1}{3}}\left|e^{\prime}_{g1}, \uparrow\right\rangle \\
&\left|\frac{1}{2}, -\frac{1}{2}\right\rangle=-\sqrt{\frac{2}{3}}\left|L_{x}=+1,\uparrow\right\rangle+\sqrt{\frac{1}{3}}\left|e^{\prime}_{g1}, \downarrow\right\rangle \\
&\left|\frac{3}{2}, \frac{3}{2}\right\rangle=\left|L_{x}=-1, \uparrow\right\rangle\\
&\left|\frac{3}{2}, -\frac{3}{2}\right\rangle=\left|L_{x}=+1, \downarrow\right\rangle\\
&\left|\frac{3}{2}, \frac{1}{2}\right\rangle=\sqrt{\frac{1}{3}}\left|L_{x}=-1,\downarrow\right\rangle+\sqrt{\frac{2}{3}}\left|e^{\prime}_{g1}, \uparrow\right\rangle \\
&\left|\frac{3}{2}, -\frac{1}{2}\right\rangle=\sqrt{\frac{1}{3}}\left|L_{x}=+1,\uparrow\right\rangle+\sqrt{\frac{2}{3}}\left|e^{\prime}_{g1}, \downarrow\right\rangle. \\
\end{aligned}
\end{equation}
The DOS results are shown in Fig. \ref{fig:7}(c). We find that the $j_{\rm eff}=\frac{1}{2}$ and $j_{\rm eff}=\frac{3}{2}$ states have significant mixtures, and this was also found in previous works\cite{Johnson_2015PRB,Yadav2016}. The mixture may also be partially due to the Ru-Ru intersite hopping and the band formation\cite{epl2016,prl2012,scirep2014}. Moreover, we test the increasing $U$ values of 3 eV and 4 eV to see a possible influence of the enhanced atomic effect (and reducing band hybridization) on the $j_{\rm eff}=\frac{1}{2}$ picture, but we find that the $S=\frac{1}{2}$ and $L_x=1$ state persists and thus the strong mixture between the $j_{\rm eff}=\frac{1}{2}$ and $j_{\rm eff}=\frac{3}{2}$ remains, as seen in Fig. S4 in the SM~\cite{SM}. Note that the $t_{2g}$ hole state $\left|L_{x}=-1,\downarrow\right\rangle$ in Fig. \ref{fig:7}(a) can be written as
\begin{equation}
\begin{aligned}
\left|L_{x}=-1,\downarrow\right\rangle=-\sqrt{\frac{2}{3}}\left|\frac{1}{2}, \frac{1}{2}\right\rangle+\sqrt{\frac{1}{3}}\left|\frac{3}{2}, \frac{1}{2}\right\rangle.
\end{aligned}
\end{equation}
Indeed, this composition of the $t_{2g}$ hole state is clearly seen in Fig. \ref{fig:7}(c), and thus the $j_{\rm eff}=\frac{1}{2}$ state seems not to be a good eigenstate. We still attempt to obtain the $j_{\rm eff}=\frac{1}{2}$ eigenstate in our LSDA+SOC+$U$ calculations by setting the corresponding occupation density matrix, however, it eventually converges to the $S=\frac{1}{2}$ and $L_x=1$ state. All these results suggest that with a considerable Hund exchange, trigonal crystal field splitting, and moderate SOC, the $j_{\rm eff}=\frac{1}{2}$ state could not be the most suitable picture for RuCl$_{3}$.

Moreover, we test other configurations for the Ru$^{3+}$ $t_{2g}^{5}$ electrons. For example, when we set the spins along the z axis, we will get the $S=\frac{1}{2}$ and $L_z=1$ state via the SOC-mixing of $e'_{g1}$ and $e'_{g2}$ orbitals. However, this state has a higher total energy than the $S=\frac{1}{2}$ and $L_x=1$ state by 26.7 meV/f.u. This accords with the experimental easy in-plane magnetization and the larger in-plane g factor than the out-of-plane one~\cite{Majumder_2015PRB,Sears_2015PRB,Johnson_2015PRB,Kubota_2016PRB}. Therefore, our results show that RuCl$_{3}$ bulk is a Mott insulator in the formal $S=\frac{1}{2}$ and $L=1$ state with the easy in-plane magnetization, and the in-plane effective moment $\mu_{\rm eff}$=$\sqrt{g^{2}_{s}S(S+1)+g^{2}_{l}L(L+1)}$$\approx$ 2.24 $\mu_{\rm B}$ could well account for the experimental one of 2.0-2.4 $\mu_{\rm B}$\cite{Majumder_2015PRB,Sears_2015PRB,banerjee_2016nm,Banerjee_2017science}. Our results provide new perspectives to understand the electronic and magnetic properties of RuCl$_{3}$.

\subsection*{D. RuCl$_{3}$ monolayer: $S=\frac{1}{2}$ and $L=1$ Mott-insulating state}

\begin{figure}[t]
  \centering
\includegraphics[width=8.5cm]{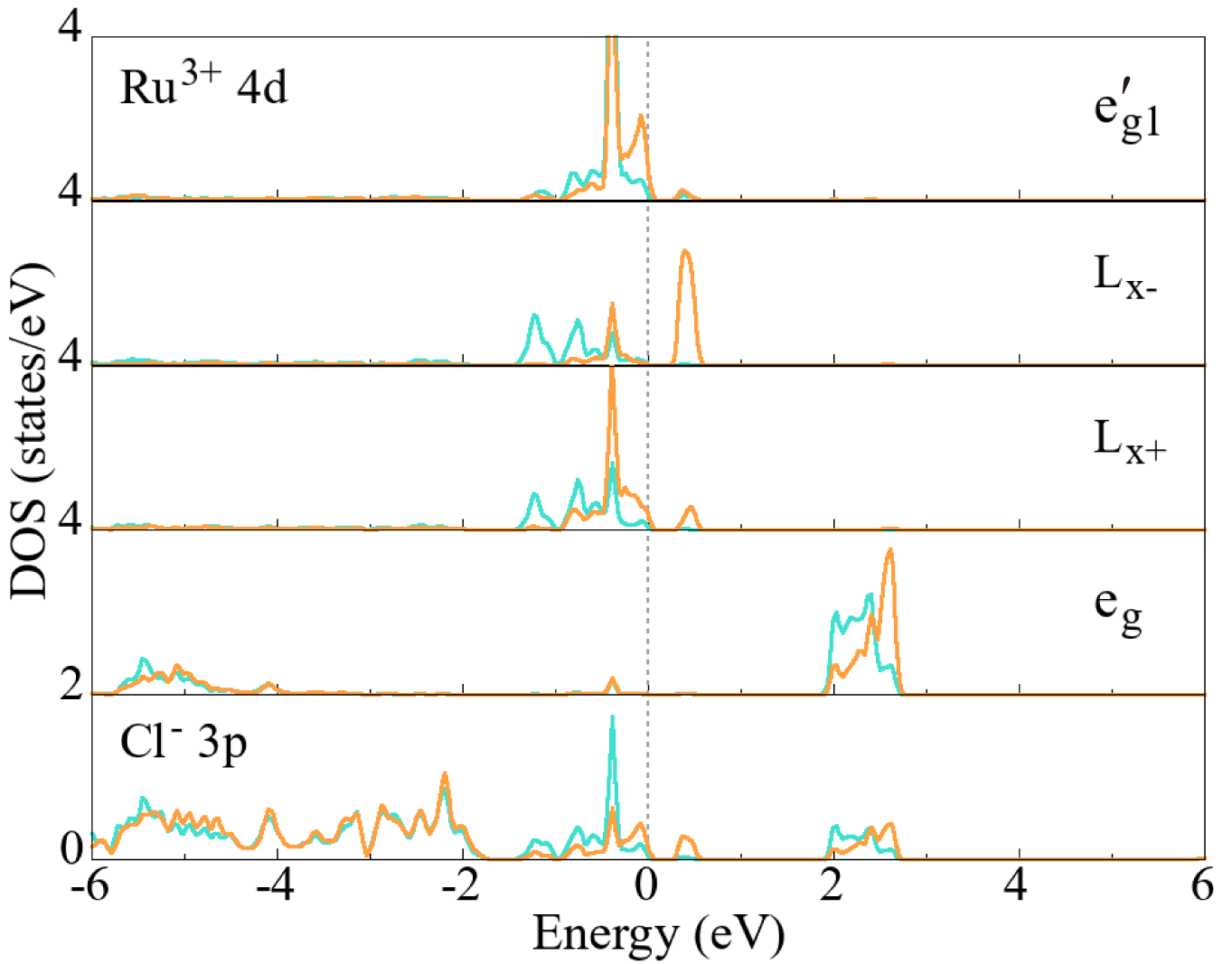}
  \caption{The DOS result of LSDA+SOC+U for RuCl$_{3}$ monolayer. The green (orange) curves refer to up (down) spins. The Fermi level is set at the zero energy.}
  \label{fig:8}
\end{figure}

The cleavage energy for RuCl$_{3}$ monolayers is calculated by the DFT+vdW correction to be 0.17 J/m$^{2}$, and it is even smaller than that of 0.24 for RuI$_{3}$ monolayers and 0.3 J/m$^{2}$ for CrI$_{3}$ monolayers. We now explore the electronic and magnetic properties of the RuCl$_{3}$ monolayer. The calculated electronic structures are very similar to RuCl$_3$ bulk and are thus not described here in a repeated way. We just stress the LSDA+SOC+$U$ results (see Fig. \ref{fig:8}), which confirm that the RuCl$_{3}$ monolayer is a Mott insulator in the formal $S=\frac{1}{2}$ and $L=1$ state with the easy in-plane magnetization. The zigzag AFM state is more stable than the nonmagnetic one by 130.8 meV/f.u., and it has a local spin moment of 0.70 $\mu_{\rm B}$/Ru and an in-plane orbital moment of 0.46 $\mu_{\rm B}$/Ru.

\section{Summary}
In light of all the above results, we find that RuI$_{3}$ and RuCl$_{3}$ display contrasting electronic and magnetic properties due to a delicate interplay of SOC, Hund's coupling, crystal field effect, transition metal-ligand hybridization, and electron correlation. When one considers the SOC effect of Ru $4d$ electrons in the local octahedral crystal field, the splitting of $j_{\rm eff}=\frac{1}{2}$ and $j_{\rm eff}=\frac{3}{2}$ orbitals would give the half-filled $j_{\rm eff}=\frac{1}{2}$ bands for the Ru$^{3+}$ t$_{2g}^{5}$ configuration. This scenario is the same for both RuI$_{3}$ and RuCl$_{3}$. However, when including the effect of Hund's coupling, we find that RuI$_{3}$ tends to be in the nonmagnetic state whereas RuCl$_{3}$ prefers to be in the spin-polarized zigzag AFM state. It is not surprising, given that for RuI$_{3}$, the lower electronegativity of the ligand iodine and its wide $5p$ orbital give rise to stronger Ru $4d$-I $5p$ hybridization, and thus to the stronger delocalization behavior and weaker spin polarization for Ru$^{3+}$ $4d$ electrons compared with those in RuCl$_{3}$. As a result, with the assistance of dominant SOC and negligible Hund's coupling, the RuI$_{3}$ bulk and monolayer are in the spin-orbital entangled $j_{\rm eff}=\frac{1}{2}$ nonmagnetic state. Moreover, aided by the strong SOC of I $5p$ orbitals and their hybridization with Ru $4d$ orbitals, the RuI$_{3}$ monolayer displays the $j_{\rm eff}=\frac{1}{2}$ band insulating behavior. In contrast, the RuCl$_{3}$ bulk and monolayer have considerable Hund's exchange splitting which is somewhat larger than the SOC strength, and combined with moderate trigonal crystal field splitting, they tend to have a large spin and in-plane orbital moment, thus approaching the formal $S=\frac{1}{2}$ and $L=1$ state rather than the $j_{\rm eff}=\frac{1}{2}$ state. Moreover, the electron correlation opens a Mott-insulating gap for more ionic RuCl$_3$ with stronger electron localization. Then, these results for RuCl$_3$ well account for the experimental effective magnetic moment, the large in-plane magnetic moment and in-plane anisotropy\cite{Majumder_2015PRB,Sears_2015PRB,Johnson_2015PRB,Kubota_2016PRB,banerjee_2016nm}. Therefore, albeit the same t$_{2g}^{5}$ configuration, the contrasting electronic structures in RuI$_{3}$ and RuCl$_{3}$ give rise to the varying electronic and magnetic properties: RuI$_{3}$ bulk is a PM bad metal and RuCl$_{3}$ bulk is a zigzag AFM Mott insulator. Our results well agree with those experiments\cite{Cava_2022am,Nawa_2021JPSJ,Majumder_2015PRB,Sears_2015PRB,Johnson_2015PRB,Kubota_2016PRB,banerjee_2016nm,Cao_2016PRB,Park_2016arxiv,Banerjee_2017science,Do_2017natphys}. Moreover, we predict a metal-insulator transition for RuI$_{3}$ from bulk to monolayer, but the RuCl$_{3}$ monolayer persists in the Mott-insulating state.

In summary, using density functional calculations, we confirm that RuI$_{3}$ bulk is a PM bad metal\cite{Cava_2022am,Nawa_2021JPSJ} on the verge of the metal-insulator transition\cite{Kaib_2022arx} and has the Ru$^{3+}$ spin-orbital entangled $j_{\rm eff}=\frac{1}{2}$ state. Our results are consistent with the experimental observations. Moreover, a metal-insulator transition occurs for RuI$_{3}$ from bulk to monolayer, and the gap opening in the RuI$_{3}$ monolayer is mainly due to the band narrowing with the decreasing lattice dimensionality and to the altered Ru $4d$-I $5p$ hybridization by the strong I $5p$ SOC effect. In contrast with the $j_{\rm eff}=\frac{1}{2}$ state in RuI$_{3}$, RuCl$_{3}$ turns out to be in the $S=\frac{1}{2}$ and $L=1$ state with a stronger in-plane magnetic anisotropy. Its Mott-insulating state and the zigzag AFM state arise from the delicate interplay of electron correlation, Hund's coupling, SOC, and trigonal crystal field distortion. These results well explain the experimental effective magnetic moment and strong in-plane magnetization of RuCl$_{3}$ bulk. We conclude that the varying electronic and magnetic properties of RuI$_{3}$ and RuCl$_{3}$ are ascribed to the contrasting electronic structures, particularly the $j_{\rm eff}=\frac{1}{2}$ state for the former, and the $S=\frac{1}{2}$ and $L=1$ state for the latter. Thus this work highlights the distinct Ru$^{3+}$ spin-orbital states and the importance of subtle interactions among various degrees of freedom.

\section*{Acknowledgements}
This work was supported by the National Natural Science Foundation
of China (Grants No. 12174062, No. 12241402, and No. 12104307).

\bibliography{arxiv} 

\setcounter{figure}{0}
\setcounter{table}{0}
\renewcommand{\thefigure}{S\arabic{figure}}
\renewcommand{\thetable}{S\arabic{table}}

\newpage

\begin{widetext}
\subsection*{\Large Supplemental Material}
\subsection*{I. RuCl$_{3}$ bulk in $P3_{1}12$ and $R\bar{3}$ structures}

In addition to the $C2/m$ structure calculated in the main text, here we also consider the $P3_{1}12$ and $R\bar{3}$ structures for RuCl$_{3}$ bulk. The lattice constants are optimized to be $a$=$b$=5.957 \AA~and $c$=17.183 \AA~for the $P3_{1}12$ structure, and a=b=5.931 \AA~and c=17.095 \AA~for the $R\bar{3}$ one. We find that the $P3_{1}12$ and $R\bar{3}$ structures give very similar results with the $C2/m$ one. The LSDA+SOC+U calculations show that the Ru$^{3+}$ ion has local spin moment of 0.70 (0.69) $\mu_{\rm B}$/Ru and an in-plane orbital moment of 0.54 (0.55) $\mu_{\rm B}$/Ru in $P3_{1}12$ ($R\bar{3}$) structure, giving the $S=\frac{1}{2}$ and $L_{x}=1$ state (see Fig. S1). The zigzag AF state is more stable than the nonmagnetic one by 124.9 (131.2) meV/fu in $P3_{1}12$ ($R\bar{3}$) structure. While these results show insignificant numerical differences among the three different structures, a same conclusion can be drawn, that is, with the assistance of Hund’s coupling, SOC, electron correlation, and trigonal crystal field, RuCl$_{3}$ bulk is in the $S=\frac{1}{2}$ and $L_{x}=1$ Mott insulating state.

\begin{figure*}[htbp]
  \centering
\includegraphics[width=15cm]{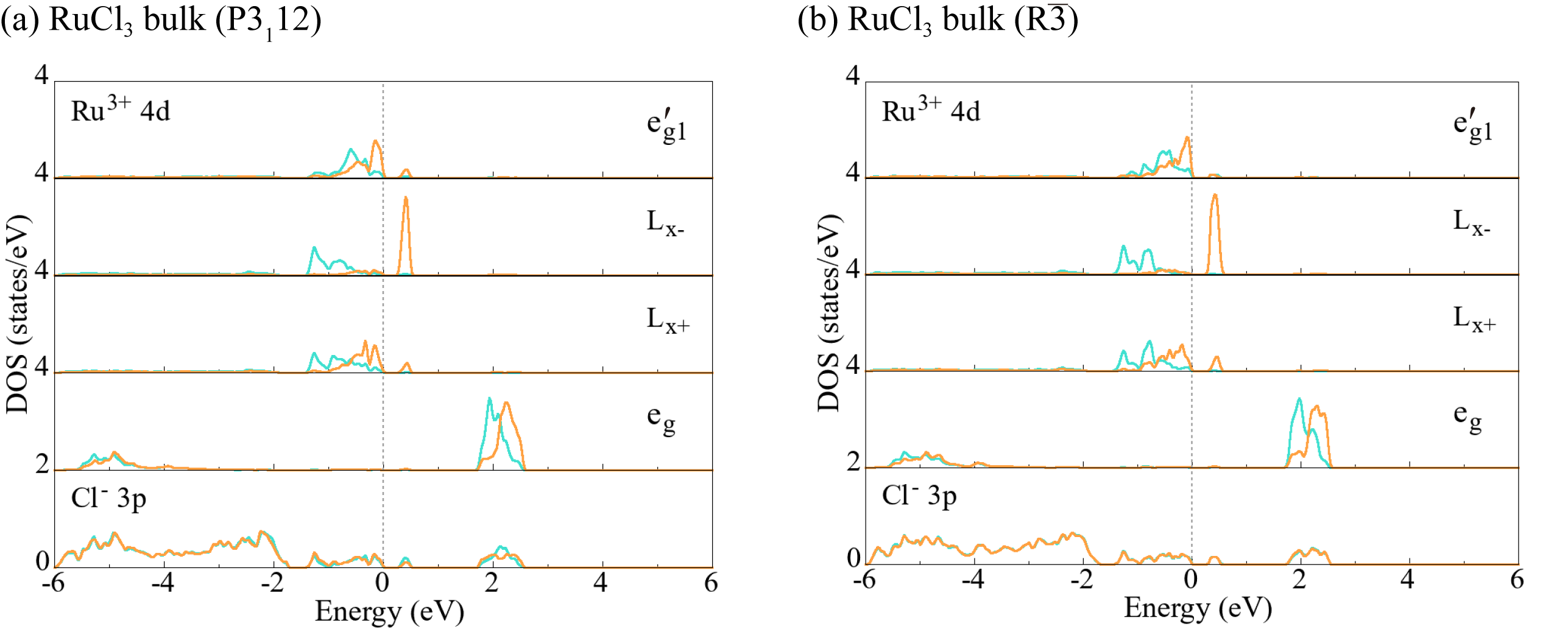}
  \caption{The DOS results of the $S=\frac{1}{2}$ and $L_{x}=1$ zigzag AF state of RuCl$_{3}$ bulk in (a) $P3_{1}12$ and (b) $R\bar{3}$ structures by LSDA+SOC+U calculations. The green (orange) curves refer to up (down) spins. The Fermi level is set at zero energy.}
  \label{fig:S4}
\end{figure*}

\newpage
\subsection*{II. LSDA+SOC+U and GGA+SOC+U for RuI$_{3}$}

\begin{figure*}[h]
  \centering
\includegraphics[width=15cm]{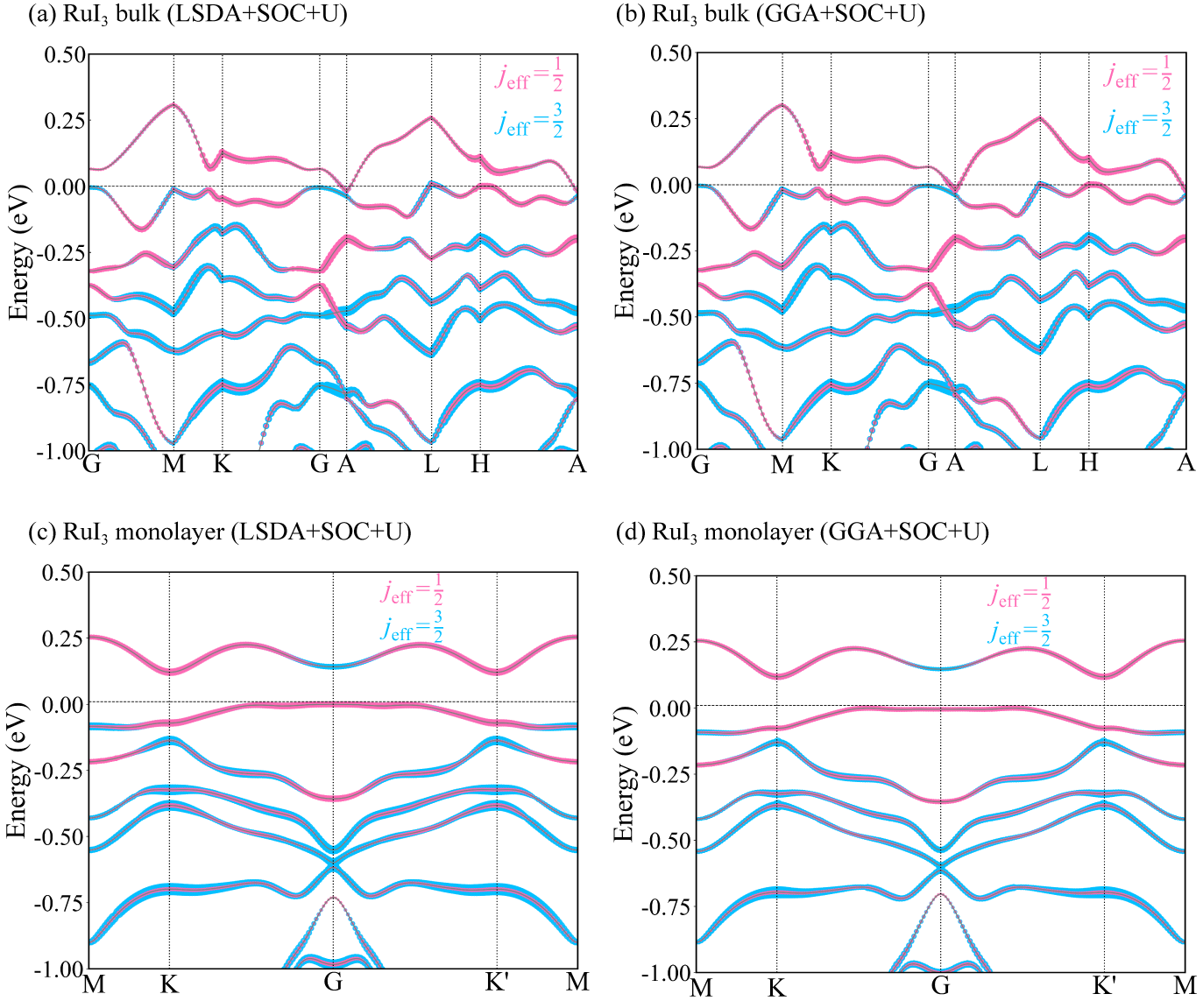}
  \caption{Band structures for the nonmagnetic state of RuI$_{3}$ bulk calculated by (a) LSDA+SOC+U and (b) GGA+SOC+U. Band structures for the nonmagnetic state of RuI$_{3}$ monolayer calculated by (c) LSDA+SOC+U and (d) GGA+SOC+U. The pink (blue) curves stand for the $j_{\rm eff}=\frac{1}{2}$ ($j_{\rm eff}=\frac{3}{2}$) bands. The Fermi level is set at the zero energy. In both LSDA+SOC+U and GGA+SOC+U calculations, the $j_{\rm eff}=\frac{1}{2}$ paramagnetic state of RuI$_{3}$ bulk and monolayer is obtained and a metal-insulator transition occurs from RuI$_{3}$ bulk to monolayer.}
  \label{fig:S1}
\end{figure*}

\newpage
\subsection*{III. LSDA+SOC+U and GGA+SOC+U for RuCl$_{3}$}
\begin{figure*}[htbp]
  \centering
\includegraphics[width=15cm]{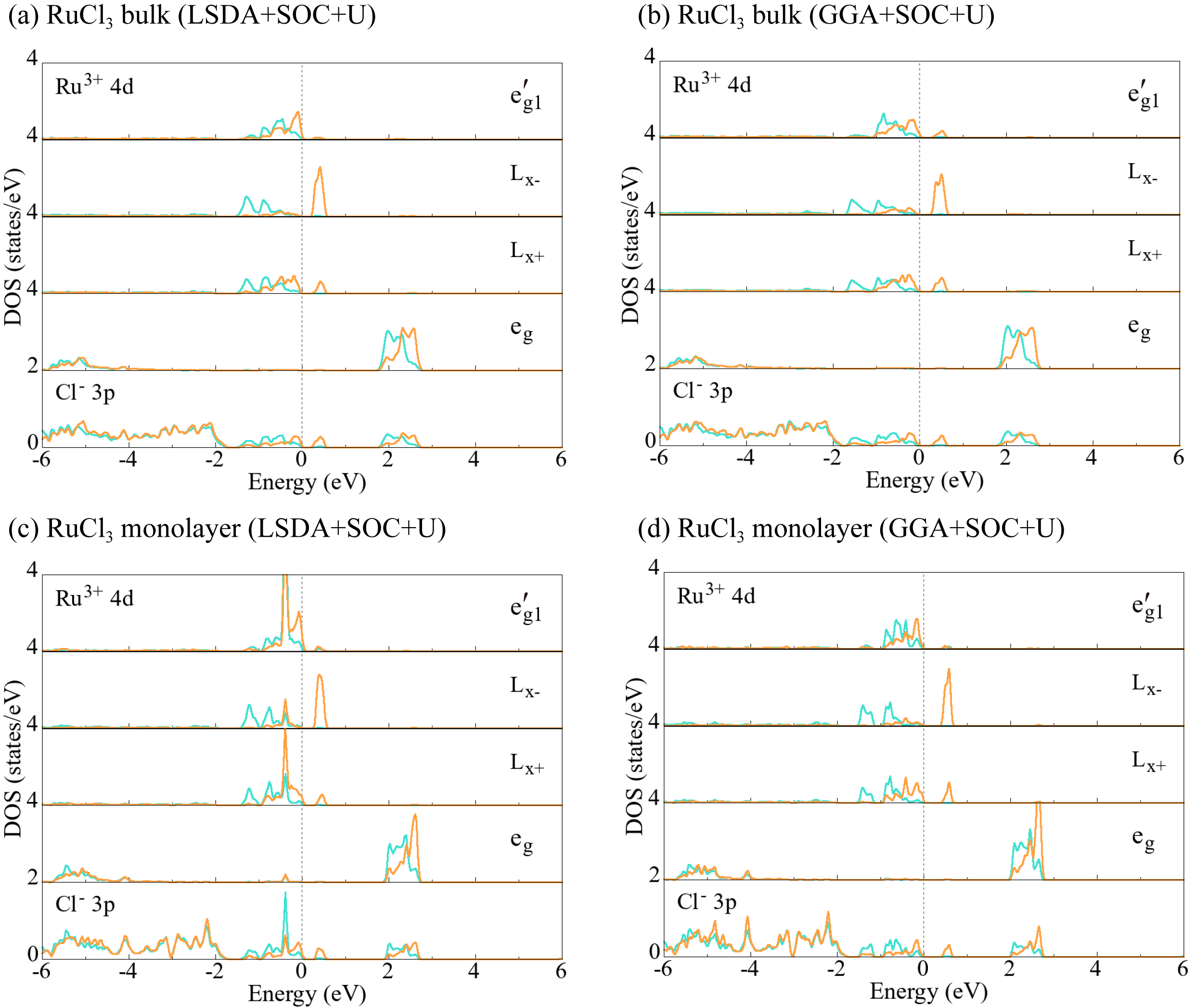}
  \caption{DOS results for the zigzag AF state of RuCl$_{3}$ bulk calculated by (a) LSDA+SOC+U and (b) GGA+SOC+U. DOS results for the zigzag AF state of RuCl$_{3}$ monolayer calculated by (c) LSDA+SOC+U and (d) GGA+SOC+U. The green (orange) curves refer to up (down) spins. The Fermi level is set at the zero energy. The local spin/orbital moment of the Ru$^{3+}$ ion in RuCl$_{3}$ bulk is calculated to be 0.70/0.47 and 0.75/0.35 $\mu_{\rm B}$/Ru by LSDA+SOC+U and GGA+SOC+U, respectively. The local spin/orbital moment of the Ru$^{3+}$ ion RuCl$_{3}$ monolayer is calculated to be 0.70/0.46 and 0.73/0.43 $\mu_{\rm B}$/Ru by LSDA+SOC+U and GGA+SOC+U, respectively.}
  \label{fig:S3}
\end{figure*}

\newpage
\subsection*{IV. larger U values for RuCl$_{3}$ bulk}
\begin{figure*}[htbp]
  \centering
\includegraphics[width=15cm]{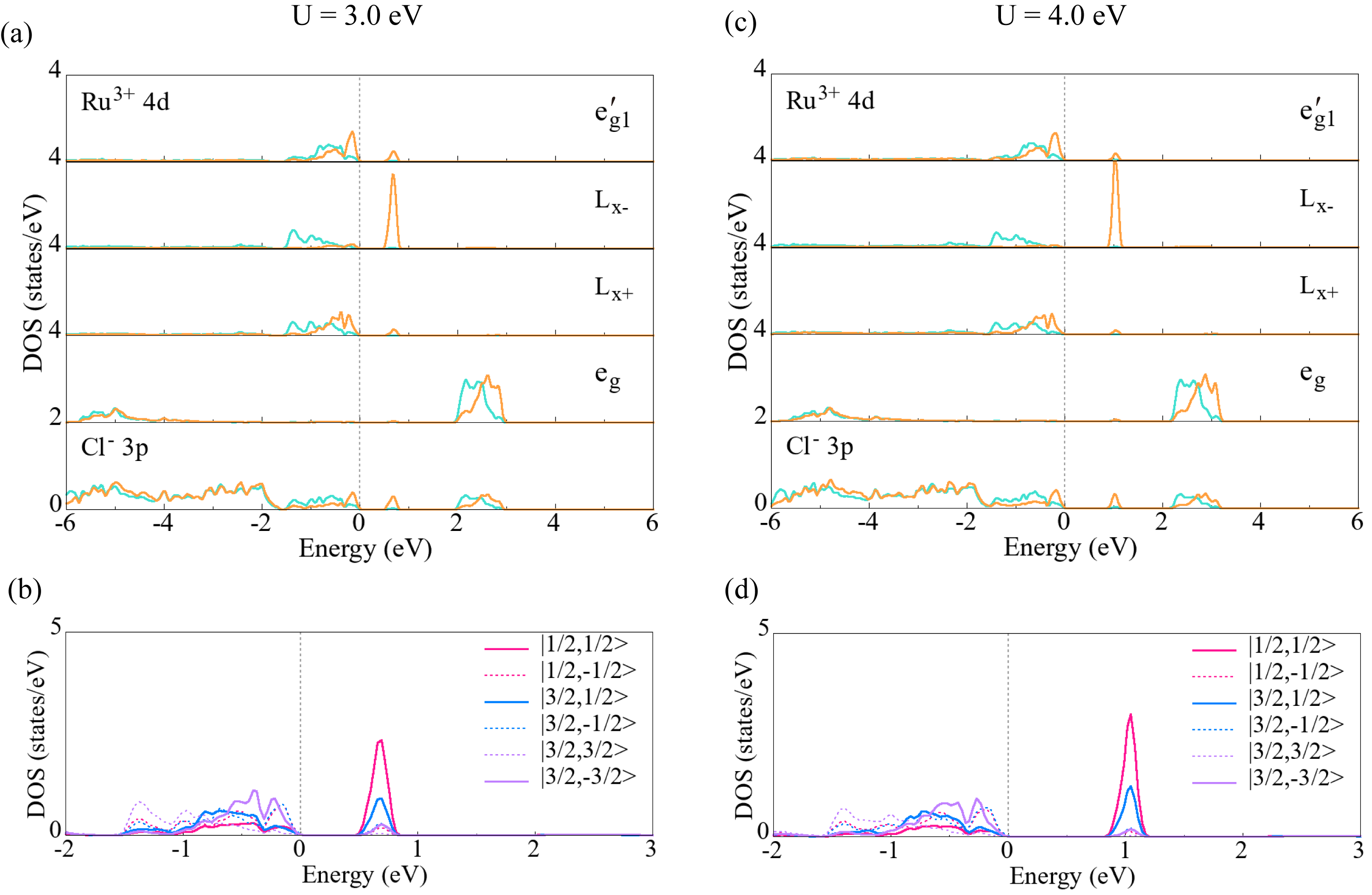}
  \caption{DOS results for the zigzag AF state of RuCl$_{3}$ bulk, calculated by LSDA+SOC+U with $J_{\rm H}=0.5$ eV and U values of (a, b) 3 eV and (c, d) 4 eV. The $t_{2g}$ states in (a) and (c) are projected onto the $\left|j_{\rm eff}, m^{x}_{j}\right\rangle$ basis in (b) and (d), respectively. The Fermi level is set at the zero energy. The DOS results closely resemble those shown in Fig. 7, except for the gap size, and all of them support the $S=\frac{1}{2}$ and $L=1$ Mott insulating state of RuCl$_{3}$.}
  \label{fig:S4}
\end{figure*}
\end{widetext}

\end{document}